\documentclass[aps,pre,preprint,amsmath,amssyyb,showpacs,twocolumn]{revtex4-2}

\usepackage{dcolumn}
\usepackage{bm}
\usepackage{color}

\usepackage{epsfig}

\usepackage{amsthm}
\usepackage{graphicx}

\usepackage{color}
\usepackage[utf8]{inputenc}
\usepackage[english]{babel}
\usepackage[titletoc,toc,title]{appendix}

\usepackage[top=2 cm,bottom=2 cm,left=2. cm, right=2. cm]{geometry}

\usepackage{dsfont}
{\nopagebreak\refstepcounter{theorem} \par{\it\ Assumption}\ \thetheorem{}. \ignorespaces\nopagebreak}%

\begin{document}


\title{Glassy dynamics near the interpolation transition in deep recurrent networks}


\author{John Hertz}
\email[]{hertz@nbi.ku.dk}


\affiliation{Nordita, Stockholm University and KTH, Sweden\\
Niels Bohr Institute, University of Copenhagen, Denmark}

\author{Joanna Tyrcha}
\email[]{joanna@math.su.se}
\affiliation{Matematiska institutionen, Stockholm University, Sweden}


\date{\today}

\begin{abstract}
We examine learning dynamics in deep recurrent networks, focusing on the behavior near the boundary in the depth-width plane separating under- from overparametrized networks, known as the interpolation transition.  The training data are Bach chorales in 4-part harmony, and the learning is by stochastic gradient descent with a cross-entropy loss function. We find critical slowing down of the learning approaching the transition from the overparametrized side:  For a given network depth, learning times to reach small training loss values appear to diverge proportional to  $1/(w - w_c)$ as the width $w$ approaches a (loss-dependent) critical value $w_c$.  We identify the zero-loss limit of this value with the interpolation transition. We also study aging (the slowing down of fluctuations as the time since the beginning of learning increases).  Taking a system that has been learning for a time $\tau_{w}$, we measure the subsequent mean-square fluctuations of the weight values at times $\tau > \tau_{w}$.  In the underparametrized phase, we find that they are well-described by a single function of $\tau/\tau_{w}$.  While this scaling holds approximately at short times at the transition and in the overparametrized phase, it breaks down at longer times when the training loss gets close to the lower limit imposed by the stochastic gradient descent dynamics.   Both this kind of aging and the critical slowing down are also found in certain spin glass models, suggesting that those models contain the most essential features of the learning dynamics.
\end{abstract}


\maketitle

\section{Introduction}
Recurrent neural networks are an attractive tool for modeling dynamical systems based on time-series data.  They contain implicitly a mechanism for storing and using information about past states of the system to generate the next state.   This property is essential, for example, in applications to music generation, where information about the distant past can be necessary to generate the correct temporal structure in the near future.  

In understanding any learning network, an essential feature is what has come to be called the interpolation threshold: the boundary in hyperparameter space separating regions where a training set can be learned (i.e., the network can successfully interpolate between all the training points) from regions where it can't be.   Some work~\cite{Geigeretal,Spigleretal} has explored aspects of this problem for standard layered networks, exploiting an analogy with jamming transitions~\cite{Silbertetal}.  More recently, Kerr Winter and Janssen~\cite{WJ} explored learning dynamics in a variety of such networks, including effects of regularization and Langevin training noise.  In some cases they found results qualitatively similar to ours, as we will discuss later.  

Of particular relevance to our work is that of Baity-Jesi {\em{et al}}~\cite{BJetal}.  Although they did not study the transition itself,  they performed interesting learning experiments for both strongly underparametrized and strongly overparametrized (non-recurrent) networks. They found spin-glass-like aging~\cite{CuDo,CrHoSo,HeShvN} in a strongly underparametrized network, as well as something similar, persisting for rather long times, in a number of strongly overparametrized networks.  However this apparent aging always came to an end as the networks approached successful learning, and the asymptotic dynamics were simple relaxation.   
 
In our work we extend this kind of study to recurrent networks, with special attention to the dynamics near the transition between underparametrized and overparameterized networks.   We choose to do this within an otherwise conventional layered network architecture trained by stochastic gradient descent (SGD), just adding full recurrence within each layer.  There are of course many possible recurrent architectures, but this seemed to be the simplest extension of the standard model network to include recurrence.  

Our aim is to gain an understanding of the asymptotic learning dynamics in both underparametrized and overparametrized networks near the transition.  In doing so, like Bati-Jesi {\em{et al}}, we will use ideas from spin glass theory, where related transitions are well-studied.  The defining characteristic of a spin glass system is the existence of ``glassy'' states where the degrees of freedom (called ``spins'') become frozen or nearly so.  The key features of spin-glass systems that are relevant to our work are the following: 

(1) Quite generally, they have phase transitions, driven by changes in temperature, external fields or other control parameters, between ``normal'' and glassy phases.  In some cases, the approach to the transition from the normal side can exhibit ``critical slowing down'', meaning that the relaxation times of the system diverge as one approaches the transition.    

(2)  The dynamical signature of the glassy phase is {\em aging}.  The basic idea of aging is this: If a system is suddenly quenched into a glassy state, it will gradually relax and approach but never reach thermal equilibrium.  If we study a system starting at a time $\tau_w$ after the quench, the properties we find will depend on $\tau_w$: the ``older'' the system (i.e., the larger $\tau_w$ is), the slower the dynamics we will observe.  In soluble model systems ~\cite{CK}, this can be expressed simply: the correlation between degrees of freedom at $\tau_w$ and at $\tau_w + \tau$, with $\tau$  of order $\tau_w$, will depend only on the ratio $\tau/\tau_w$, reflecting the fact that $\tau_w$ is the only relevant characteristic time in the system.

Further details of the spin glass models relevant to our work can be found in Appendix A.

In the present work, we have trained recurrent nets to imitate Bach's chorales (sequences of chords in 4-part harmony) and, like Baity-Jesi {\em{et al}}, asked whether and to what degree they exhibited dynamical properties like spin glass systems.  We have studied the phase diagram of these networks in the space defined by their width $w$ and depth $d$ and investigated the neighborhood of the interpolation transition as follows: For each depth considered, we find the smallest width, $w_c(l)$,  for which the loss can be trained down to a specific small target value $l$. The values of $w_c(l)$ for different $l$ form a line in the $w$-$l$ plane. We find that as one approaches this line from the $w > w_c(l)$ side, critical slowing down occurs.  The learning time to reach $w_c(l)$ diverges, and for $w < w_c(l)$ the loss can never be brought below $l$.  We identify the zero-$l$ limit of this line with the interpolation transition.  We call the region of the $w$-$l$ plane to the left of this line underparametrized and the region to the right overparametrized.   In the underparametrized region, we find aging like that in a class of mean-field spin-glass models~\cite{CuDo,CrHoSo,HeShvN}.    In the overparametrized region, this aging disappears as the learning approaches its final steady state.

The paper is organized as follows.  In Sec.~II we describe the model networks we study, the training data, and how they are trained.  In Sec.~III we present general features of the training results: the shapes of the learning curves for under-, over- and critically parametrized networks, showing how they depend on network depth.  We also examine the asymptotic long-time decay of the loss function.
Sec.~IV then explains how we observe critical slowing down and, for depths 1, 2 and 5, estimate
the critical layer size $w_c(l)$ for a range of small loss values $l$.  Sec.~V is about aging.  We consider a network that has been learning for a time $\tau_{w}$ and then measure, at longer times, the mean-square fluctuations in the connection weight values relative to their values at $\tau_{w}$.  Finally, Sec.~VI is devoted to discussion of the results, particularly with respect to theoretical findings for spin-glass models, and general conclusions.

\section{Models and training}
\label{modeltrain}

Our networks have feed-forward connections between layers and fully recurrent connections within each layer. In this architecture, the recurrence within layers allows the hidden units to store information about past time steps and thereby better predict the next step.  The multilayer structure then permits this advantage to be exploited in different ways in different layers, giving the model the potential to describe more complex long-time dynamics than a single-layer recurrent model could.     

Letting the hidden layer index $i$  run from $0$ to $d-1$, we denote the feedforward connection matrix to each layer $i$ by ${\sf J}_i$ and the recurrent connection matrix within each layer by ${\sf M}_i$.   (We do not find it necessary to use biases except for the final output.)  We have studied networks of depth $1$, $2$, $5$, and $10$ with widths ranging from $25$ to $200$.   

\begin{figure}
  \centering
  \includegraphics[width=0.46\textwidth]{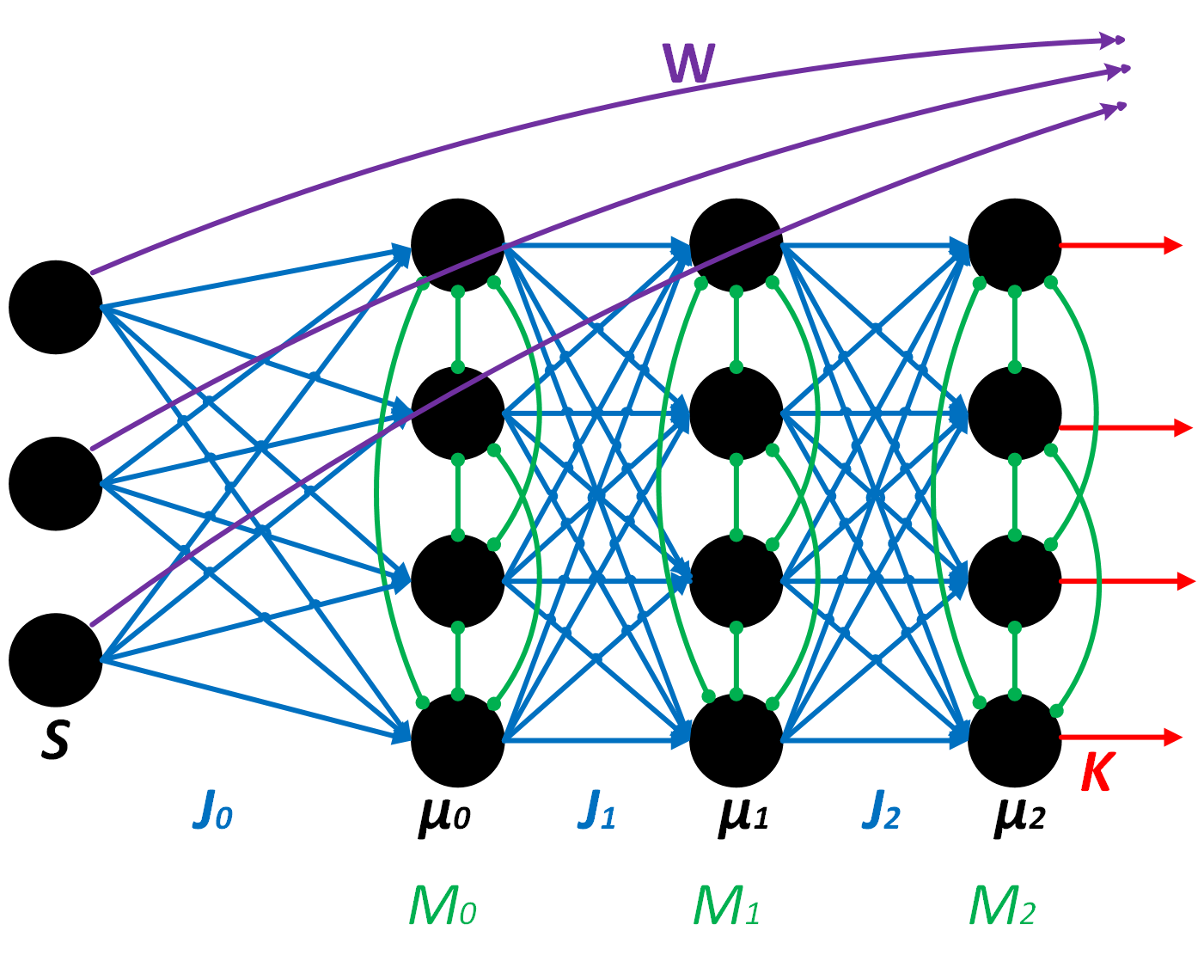} 
  \caption{Schematic graph of our network architecture. The blue lines represent feedforward connections, and each green line represents a pair of (generally unequal) intralayer interactions $J_{l,ab}$ and $J_{l,ba}$.  The purple 
(${\sf W}$) 
and red 
(${\sf K}$) 
connections feed into the output units as described by Eq.~3.}
\end{figure}

For the learning data, we use the Bach Chorale data set studied by Boulanger-Lewandowski {\em{et al}}~\cite{BL}, as described in Appendix B.   The raw 54-dimensional input vectors ${\vec S}_0(t)$ in the training set represent the chords in the chorales, and each example is a transition from one chord to the succeeding one. In running the network the input components $S_{0i}(t)$ are rescaled so that they have zero mean and unit variance. The first layer-to-layer connection matrix, ${\sf J}_0$, is $w \times 54$.  The rest of the $\sf J$ matrices are $w \times w$, as are the intralayer recurrent interaction matrices $\sf M$.   There are two matrices that connect to the output units.  One of them, $\sf W$ ($54 \times 54$), connects directly from the inputs ${\vec S}_0$, and the other, denoted $\sf K$, ($54 \times w$) connects from the final hidden layer.   Finally, there is a 54-dimensional output bias vector ${\vec h}_0$. Every unit has a tanh activation function.  The network is pictured schematically in Fig.~1.

Thus, with ${\vec \mu}_l$ denoting the activation vector of layer $l$, the hidden-layer units obey the dynamics
\begin{equation}
{\vec \mu}_0(t) = \tanh [{\sf J}_0 {\vec S}(t) +{\sf M}_0{\vec \mu}_0(t-1)]    			\label{hidden0}  
\end{equation}
and
\begin{equation}
{\vec \mu}_l(t) = \tanh [{\sf J}_l {\vec \mu}_{l-1}(t) +{\sf M}_l{\vec \mu}_l(t-1)]   		\label{hiddenl}
\end{equation}
for $1 \le l \le d-1$.  We consider the output units $\sigma_i(t)$ as stochastic, taking on the values $\pm1$ with probabilities $\exp (h_i(t)\sigma_i(t))/(2 \cosh h_i(t))$, where $h_i(t)$ are the components of
\begin{equation}
{\vec h}(t) = {\sf K} {\vec \mu}_{d-1,}(t) + {\sf W} {\vec S}_0(t) +{\vec h}_0 . 		\label{dynamics}
\end{equation}

We trained our networks using back propagation through time (BPTT) and stochastic gradient descent (SGD) with a minibatch size of 300 chord transitions and a cross-entropy loss function.   In training, the next chord in the chorale, $S_0(t+1)$, is used as the output target at step $t$. When the trained network is used to generate new music, the output at one time step is fed back as the input at the next step.    

In order to avoid exploding or vanishing gradients in the learning, we initialized all our matrices ${\sf J}_i$ (except $i=0$) and ${\sf M}_i$ as orthogonal, as was done for linear non-recurrent feedforward networks by Saxe {\em{et al}}~\cite{Saxeetal} and extended to layered networks with general activation functions by 
Pennington {\em{et al}}~\cite{Penningtonetal}.    Here, we found that stability required us to to enforce continued orthogonality of the ${\sf M}_i$ (though not the ${\sf J}_i$) under learning.    

The training process is described in greater detail in Appendix C.

\section{General features of the learning}
\label{gen_features}

We trained networks of depth $d =$ 1, 2, and 5. Final loss values were always 0.01 or smaller.  (Here and henceforth whenever we write ``loss'' we will mean its value in bits per output unit per time step.)  The $0.01$-bit value corresponded to an accuracy of approximately $99.9\%$.   

Panels (a), (b), and (c) of Fig.~2 show typical learning curves for the networks of three depths.  In each graph, we show cases where the network is underparametrized, overparametrized, and somewhere near the critical value for that depth.  The learning time variable $\tau$  is defined by
\begin{equation}
\tau(n) = \sum_{m<n} \eta(m),
\end{equation}
where $n$ and $m$ count the number of iterations and $\eta(m)$ is the learning rate at step $m$.  We sometimes refer to $\tau$ as {\em{proper time}}.
 
\begin{figure}
  \centering
(a) \\
  \includegraphics[width=0.4\textwidth]{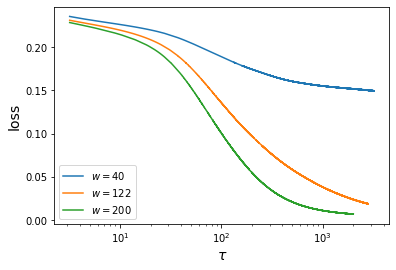} \\
(b) \\
  \includegraphics[width=0.4\textwidth]{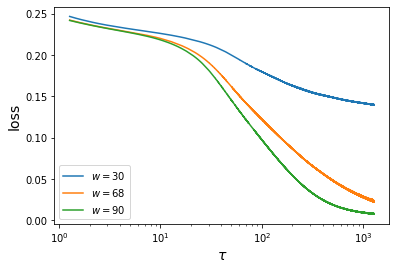} \\
(c) \\
  \includegraphics[width=0.4\textwidth]{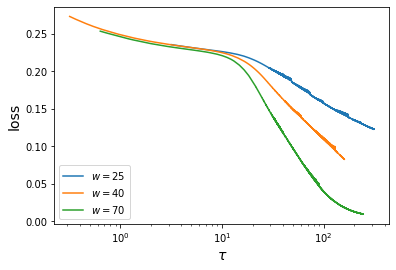} \\
(d) \\ 
  \includegraphics[width=0.4\textwidth]{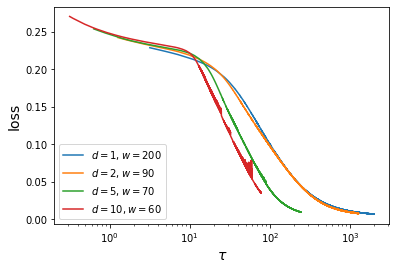}
  \caption{Sample learning curves for (a) $d= 1$, (b) $d = 2$, and (c) $d = 5$. (d):  Learning curves for overparametrized cases for  depths $d = 1$, $2$, $5$, and $10$.  The scale for the learning time $\tau$ is logarithmic.}
\end{figure}

Increasing the network depth shrinks the learning time as measured by the variable $\tau$.  Panel (d) of Fig.~2 shows this effect (though there is apparently no significant difference between $d=1$ and $d=2$).  

We found for $d>2$ that it was necessary to reduce the learning rate $\eta$ strongly during learning in order to maintain smooth learning curves, and that this constraint grows more severe with increasing depth.  (The curve in Fig.~2 for $d=10$ illustrates the kind of noise that can occur if the learning rate is not reduced strongly enough.)  This effect, coupled with the fact that deeper networks require more calculations to begin with, means that the actual computation time grows rather than shrinking with increasing depth.   

For underparametrized cases, we see that, as expected, the loss remains far from zero even for the longest learning times.  In the neighborhood of the critical layer size, it is evident that the learning becomes very slow.  For the overparametrized cases, the loss converges toward very small values, much more quickly than at criticality. 

In some cases, we also calculated the loss on test data.  In all such cases, we find that the test losses reached minima early in the learning and rose smoothly thereafter.  These results are shown in Appendix D.

\subsection{Asymptotic loss decay}

For $d=2$ we were able to make very long runs, up to $\tau \approx 5000$ and explore the asymptotic decay of the loss for the 3 values of $w$ for which learning curves were shown in Fig.~2b.  For all three cases, we found power-law dependence on the learning time $\tau$, as shown in Fig.~3.  For the underparametrized case $w=30$, we found power-law decay to an asymptotic loss value $L_\infty  = 0.1265$ with an exponent $0.52$.  For the very-nearly critical width value $w_c=68$ we found a very small $L_\infty$ of $0.002$ and an exponent $0.83$, and, for the overparametrized case $w=90$, an $L_\infty$ of $0.0035$ and an exponent $1.16$.   The non-zero asymptotic losses in  these cases are consistent with the fact that, for SGD learning, the loss can never reach zero because of the finite minibatch size. 

The exponents in these fits are slightly sensitive to the assumed asymptotic loss value; we found that changes in the asymptotic loss of size 0.0015 could lead to as large as 4\% changes in the inferred exponent, with no visible change in the fit quality.  For this reason, the given estimated exponents are correspondingly uncertain, and the estimated asympotic loss values for the critical and overparametrized cases can not be considered accurate (though we know they have to be positive).   

These results were all obtained from single learning runs.  Because these runs took so long, we were not able to make a systematic exploration of how the results might vary across different realizations of the learning process.  However, we did repeat the fit for a $w=30$ network with a different random initialization of the weights and found no difference in the estimated decay exponent (within the uncertainty level discussed about).  While more extensive exploration is possibly desirable, we do not expect that the exponents are likely to show significant variation across initial weight value choices or across learning runs.
 
\begin{figure}
  \centering
  \includegraphics[width=0.48\textwidth]{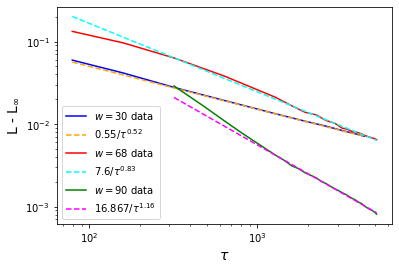} 
  \caption{Asymptotic power-law approach of the loss to asymptotic values $L_\infty$ for $d=2$, $w = 30$, $68$, and $90$.  Loss values were measured every $4 \times 10^5$ learning steps.  At the longest times $(\tau > 2\times 10^3$) the measurements were averaged over 1000 steps to reduce the effect of SGD noise.  The dashed orange lines are the power-law fits.}
\end{figure}

For $d = 1$ and $5$, we have fewer data, but again for the overparametrized cases we found them consistent with power-law decay and made estimates of the decay exponents in the same way.  For $d=1$ and $w=200$, we found an exponent  $1.04$, and for $d=5$  and $w=70$ we found $1.57$, with at least the same level of uncertainty as for the $w=2$ cases.  Table 1 shows all our results for these exponents. The overall pattern suggests that the decay exponent increases with both $d$ and $w$.  However, the detailed dependence on $d$ and $w$ remains to be explored.  

\begin{table}[h]
\caption{Asymptotic decay exponents}
\begin{tabular}{c c c c c}\hline
$w$ & &$ d=1$ & $d=2$ & $d=5$ \\ \hline
30  & & & 0.52 & \\
68  & & & 0.83 &  \\
70  & &  & & 1.57 \\
90  & &  & 1.16 &  \\
200 & & 1.04 & &  \\
\hline 
\end{tabular}
\end{table}

Because of the stochasticity of SGD learning, our learning curves are always intrinsically noisy. The noise is not visible at the resolution of the plots in Fig.~2, but at the very small losses achieved at very long times, a closer look at the data reveals fluctuations that are not negligible compared to the time-averaged mean values. Fig.~4 illustrates this fact in a 1000-step graph of the measured loss after $3.2 \times 10^6$ learning steps for $d=2$ and $w=90$, for minibatch sizes $p= 75$ and $300$.  As expected, smaller batches lead to greater noise and larger mean error.  We chose a larger-than-usual batch size of 300 in our calculations (except this one) to reduce this effect.  Nevertheless, we still had to use 1000-step mean values of the measured losses at the longest times to get the smooth results seen in Fig.~3. 

\begin{figure}
 \centering
  \includegraphics[width=0.48\textwidth]{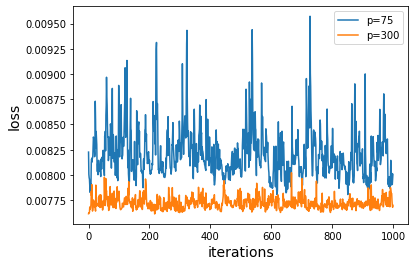}
  \caption{Fluctuating losses after $3.2 \times 10^6$ iterations for SGD minibatch sizes 75 and 300.}
\end{figure}

\section{Critical dynamics} 
\label{transition}

In order to make estimates of the critical values $w_c(l)$ of the number $w$ of hidden units per layer for given depth and target loss $l$, we studied systems with $d=$ $1$, $2$ and $5$.  For each $d$ we trained the networks for a range of values of $w$.  We recorded the proper times required for the loss to reach target values $l =$ $0.01$, $0.02$, $0.03$, $0.05$ and $0.1$.  All these times grew dramatically as $w$ was reduced.  Plotting the reciprocals of these learning times as functions of $w$ revealed that they all varied nearly linearly with $w$.  These measurements are shown as the black dots in Fig.~5.

\begin{figure}
  \centering
 (a) \\ \includegraphics[width=0.48\textwidth]{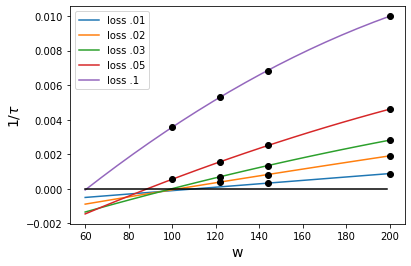} \\
 (b) \\ \includegraphics[width=0.48\textwidth]{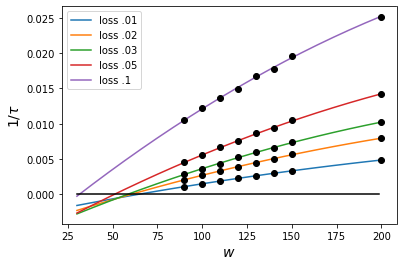} \\
  (c) \\ \includegraphics[width=0.48\textwidth]{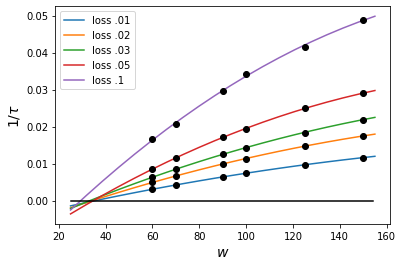}
  \caption{Inverse learning times $1/\tau$  required to reach to reach loss values  $0.01$, $0.02$, $0.03$, $0.05$ and $0.1$, as functions of $w$ for (a) $d= 1$, (b) $d = 2$, and (c) $d = 5$.  Black dots mark the results of the learning runs, and the colored curves are quadratic regression fits to them. The points where these curves cross the $w$ axis are our estimates of $w_c$.}
\end{figure}  

We made quadratic regression fits to each set of learning-time measurements.  These are shown as the colored curves in Fig.~5.  The place where the curve for a loss value $l$ crosses the $w$ axis is our estimate of \textcolor{blue}{$w_c(l)$ for that $l$}.  The graphs show that, assuming the validity of the extrapolation, the system exhibits critical slowing down as one approaches the critical value of $w$ for that loss value.  

\begin{figure}
\centering
\includegraphics[width=0.48\textwidth]{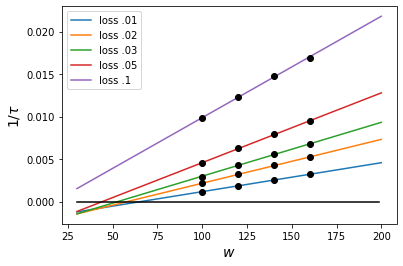}
\caption{Inverse learning times to reach specified loss values, $d=2$, with a different set of 80 chorales used for training.}
\end{figure}

For $d=2$ we also conducted the same procedure training on a different set of 100 chorales. The results are shown in Fig.~6.   While the exact values of the axis crossings were slightly different, the near-linear dependence of the inverse training times on $w$ was repeated, giving us some confidence in the robustness of the phenomenon of critical slowing down.  

We also remark that, having been obtained from different runs of different networks, the data points in Figs.~5  and 6 would not have fit the smooth curves well if run-to-run variations in the loss were significant.   This strengthens our provisional conclusion in the preceding section that such variation is can be neglected in our investigations.

As mentioned in the Introduction, critical dynamics like this also  occur in spin-glass models near the lines in a phase diagram that mark the boundary of a spin glass phase.  Notably, the Sherrington-Kirkpatrick model~\cite{SK}, arguably the simplest spin glass, has such a line in its temperature-field phase diagram~\cite{AT}.  Similarly, so does the simple perceptron near its capacity limit~\cite{GD,GR97}.   Other systems with critical slowing down near such lines in their phase diagrams include p-spin glasses in an external field~\cite{CrHoSo} or in the presence of a ferromagnetic interaction~\cite{HeShvN}.  And, finally, the problem of a particle moving in a correlated random potential in a high-dimensional space~\cite{CuDo} also turns out to be relevant.  Although the connection is not immediately apparent, it is closely related to them mathematically.  

\begin{figure}
  \centering
  (a) \\ \includegraphics[width=0.48\textwidth]{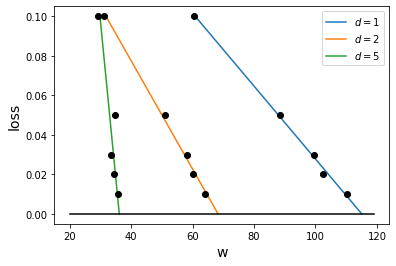} \\
   (b) \\ \includegraphics[width=0.48\textwidth]{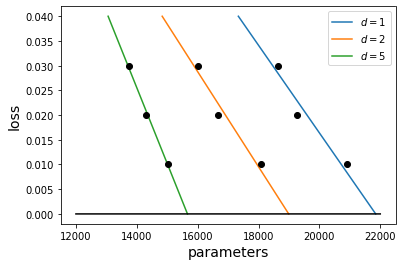}
  \caption{(a): loss vs width for $d=$ $1$, $2$ and $5$. (b) loss vs number of network parameters for $d=$ $1$, $2$ and $5$}
\end{figure}

For each depth, we performed linear regression of the estimated critical values $w_c(l)$ against the respective loss values $l$.  Fig.~7a shows the results, plotting $w_c$ on the x axis and the target loss values $l$ on the y axis.  These lines are estimates of the way the critical loss varies with $w$.  At least for $d=1$ and $2$, the linear dependence assumed in the regression appears to be reasonable, though for $d=5$ we cannot draw any immediate conclusion. We have to be cautious here because the data points in this graph are themselves results of regression and extrapolation on the data, but it at least appears clear that the deeper the network, the sharper the transition is.  

We also considered the relation between the critical loss and the total number of parameters in the model instead of $w$.  Here, we found approximate linear fits only for losses $\le 0.03$.  These are shown in Fig.~7b. It is evident that the slopes of the lines for different network depths are not as different as in Fig.~7a.  One might consider the hypothesis that they should be the same, meaning that reducing the loss reached by a given amount required a certain number of new parameters, independent of depth. However, these limited data do not support that hypothesis quantitatively.

One thing that Fig.~7b makes immediately clear, however, is that the deeper the network, the fewer weights are required to make it possible to reach a given (small) loss value.  Quantitatively, if we compare the number of parameters in the three $l$=$0.01$ critical networks, we find that the single-hidden layer one ($w_c = 110.21$) has $20890$, the two-hidden-layer one  ($w_c = 64.192$) has $18080$, and the 5-hidden-layer one ($w_c = 35.735$) has $15040$ parameters. (Here, in counting parameters, we assign the $\sf M$ matrices a number $w(w-1)/2$, where the division by 2 is because they are constrained to be orthogonal.)  Thus, depth is apparently more effective than width, at least for these data and this class of networks.

We tentatively identify the extrapolation of the lines in Fig.~7 to $l=0$ as marking the interpolation transition -- the critical values, for these depths, of the network widths required to learn the training data perfectly.  However, this identification requires a bit of explanation. Even if we could perform arbitrarily long training runs, we could not in principle obtain data points completely filling in the lines in Fig.~7  for $l$ between $0$ and $0.01$.  This is because with SGD the loss can never get all the way to zero.  In our exploration of the long-time asymptotic learning curves in the preceding section, we made a rough estimate of the minimum loss for the critical $w$ of $0.002$.  While this estimate is a bit rough, it is not negligibly smaller than $0.01$, raising the question of whether the measurements for loss 0.01 are distorted by the proximity to the asymptotic loss limit.  

To gain some insight into this question, we performed training runs like those used to make Fig.~5b with a quadratic hinge loss function and ordinary (not stochastic) gradient descent dynamics.  As Fig.~8 shows, we observe exactly the same phenomenology as with SGD, at least for losses $\ge 0.01$.  With this loss function and these noiseless training dynamics there can be no effect of a minimum loss, so we conclude that the results shown in Fig.~6 are not significantly influenced by the nonzero asymptotic loss inherent in the SGD calculation.  The extrapolations in Fig.~7 to zero loss can be thought of as what we would expect to find with an infinite minibatch size, if that were possible.

\begin{figure}
 \centering
  \includegraphics[width=0.48\textwidth]{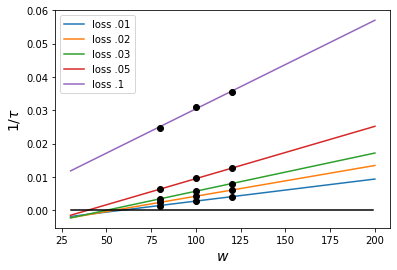}
  \caption{Inverse learning times $1/\tau$  required to reach to reach loss values  $0.01$, $0.02$, $0.03$, $0.05$ and $0.1$, as functions of $w$ for $d = 2$, for quadratic hinge loss function.  The lines are linear regression fits.}
\end{figure}

We also note that if we had chosen too small a minibatch size, the asymptotic losses would be larger than .01, and we would have seen a difference between the SGD and hinge loss results at this loss value.  If we guess that the asymptotic loss should be inversely proportional to the square root of the minibatch size, this would happen at a minibatch size $\approx 12$.

Of course, we still lack good estimates of $w_c(l)$ for $l < 0.01$ ~(also for the hinge loss function).  We do not see any reason to expect that the shape of the $w_c(l)$ curve should change dramatically in this region, so we believe that our linear extrapolation of the results from $l > 0.01$ is reasonable.

\section{Aging}
 
A pervasive feature of glassy systems is aging, i.e., the slowing down of fluctuations over long timescales~\cite{CK}.  Baity-Jesi {\em{et al}}~\cite{BJetal} explored this phenomenon in some standard layered feed-forward networks.   Their networks did not have recurrent dynamics, and they were all either strongly overparameterized or (in a single case) strongly underparametrized.  Here we study the transition region for our networks, using the same approach that they took.  The procedure was as follows:  We made long learning runs of up to $3.2 \times 10^6$ iterations, saving the network weight values after $2 \times 10^5$,  $4 \times 10^5$, $8 \times 10^5$, $1.6 \times 10^6$ and $3.2 \times 10^6$ iterations.  These run durations, converted to proper time $\tau$, are conventionally called ``waiting times'', and we denote them by $\tau_{w}$.   Then, taking each of these saved networks as an initial condition, we measured the growth of the mean-square weight fluctuations 
\begin{equation}
D(\tau_{w}, \tau_{w} +\tau) = \langle [w_{ij}(\tau_{w}+\tau) - w_{ij}(\tau_{w})]^2 \rangle_{ij}	,		\label{eq:D}
\end{equation}
as the learning was run for an additional time $\tau$.  The maximum $\tau$ was always at least three times the length $\tau_{w}$ of the initial run (and greater than that for the smaller values of $\tau_{w}$).   

To quantify how the growth of the subsequent fluctuations with $\tau$ depends on the waiting time $\tau_{w}$ we measured the increase in the mean square weight fluctuations at each increase of $\tau$ relative to the intrinsic weight fluctuations inherent in SGD at that time.  At time $\tau$, these fluctuations
have a mean-square value (a kind of temperature)
\begin{equation}
\Gamma(\tau) = \frac{1}{p}\left \langle {\rm var}_{\mu}\left( \frac{\partial L_\mu(\tau)}{\partial w_{ij}}\right) 
\right \rangle_{ij}  .														\label{eq:Gamma}
\end{equation}
Here $L_{\mu}$ is the loss function evaluated on training example $\mu$, $w_{ij}$ denotes a weight in the network, $p$ is the minibatch size, and the variance is over the minibatch of training examples for each weight.   Thus, we should evaluate the quantity
\begin{equation}
\Delta(\tau_{w}, \tau_{w} + \tau) = \int_{\tau_{w}}^{\tau_{w} + \tau} d\tau'\frac{\partial D(\tau_{w},\tau_{w}+\tau')/\partial \tau'}{\Gamma(\tau')} 											\label{eq:Delta}
\end{equation}
where the integral notation really means a sum over individual learning steps.  Calculating $\Delta$ this way, rather than just taking $\Delta \propto D$, can be thought of as removing the effect of the varying noise level $\Gamma$ produced by SGD. This permits comparison with ordinary models with constant noise (constant temperature).  

 In practice, it was computationally too costly to evaluate $\Gamma(\tau')$ at every iteration.  However, we found that $\Gamma$ varies rather slowly with $\tau$, so we only evaluated it every 10000 steps.   Baity-Jesi {\em{et al}} employed the approximation $\Gamma(\tau) = \Gamma(\tau_{w})$, which we find to be reasonable  at very large $\tau$ for overparametrized models but not for all the cases we studied.

\begin{figure*}
  \centering
(a)  \hspace{0.3\textwidth} (b) \hspace{0.3\textwidth} (c)\\   
  \includegraphics[width=0.32\textwidth]{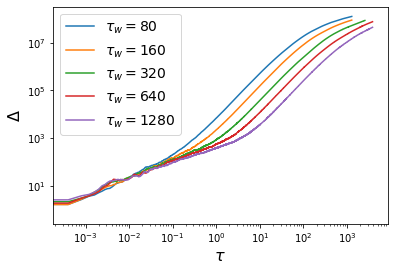} \includegraphics[width=0.32\textwidth]{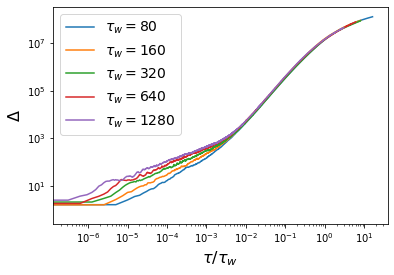} \includegraphics[width=0.32\textwidth]{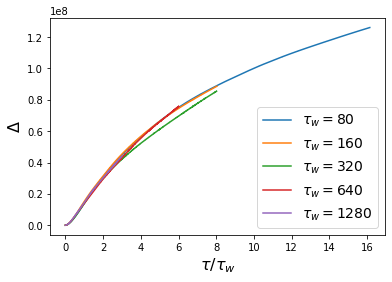}\\
  (d)  \hspace{0.3\textwidth} (e) \hspace{0.3\textwidth} (f)\\   
  \includegraphics[width=0.32\textwidth]{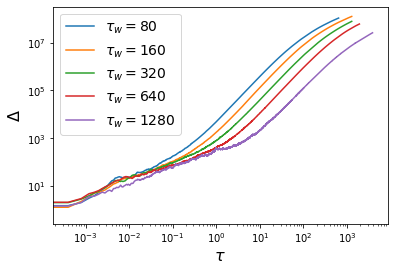}  \includegraphics[width=0.32\textwidth]{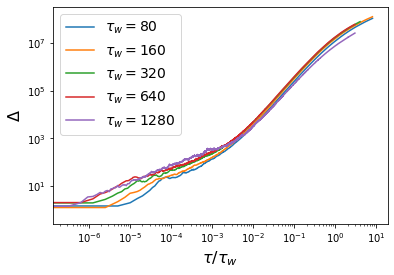}  \includegraphics[width=0.32\textwidth]{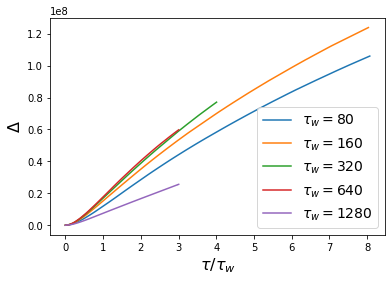} \\
  (g) \hspace{0.3\textwidth} (h) \hspace{0.3\textwidth} (i)\\   
  \includegraphics[width=0.32\textwidth]{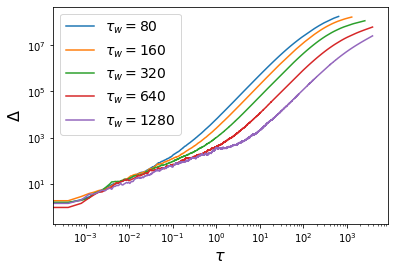}  \includegraphics[width=0.32\textwidth]{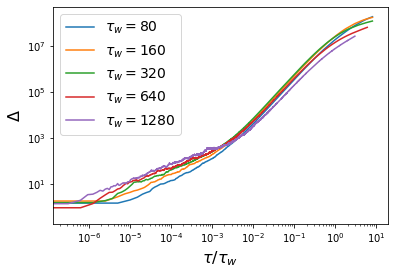} \includegraphics[width=0.32\textwidth]{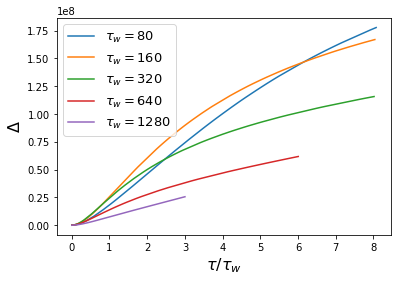} \\

  \caption{Aging for $d= 2$, $\tau_{w} = 80$, $160$, $320$, $640$, and $1280$. (a) $w=30$, loglog plot of
  $\Delta$ as function of $\tau$, (b): as in (a) but $\Delta$ plotted as function of $\tau/\tau_{w}$, (c): like (b) but linear plot; (d), (e) and (f): like (a), (b) and (c) but for $w=68$; (g), (h) and (i): like (a), (b), and (c) but for $w=90$. }
\end{figure*}

\begin{figure}
 \centering
 \includegraphics[width=0.48\textwidth]{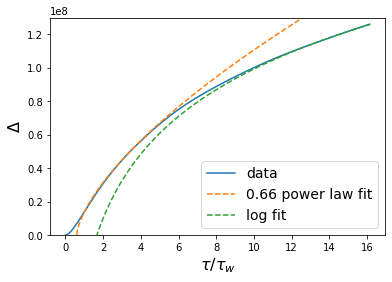}
\caption{Fits to $\Delta(\tau/\tau_{\rm w})$ for small (1-6) and large (10-16) values of $\tau/\tau_{\rm w}$.}
\end{figure}

We have performed calculations for networks with two hidden layers, with widths $30$, $68$ and $90$, to illustrate the phenomenon for moderately underparameterized, nearly critical, and moderately overparametrized cases, respectively. Following the example of Baity-Jesi {\em{et al}}~\cite{BJetal}, we show the results as log-log plots in the first two columns of Fig.~9. We also show the corresponding linear plots of $\Delta(\tau/\tau_{w})$ in the third column.

Two universal features of these plots are immediately evident.  The first (which can be seen in all the graphs in the left-hand column), is that the plots for different values of $\tau_{w}$ collapse onto a common function of $\tau$ for very short times $\tau \ll \tau_{w}$.  This effect is a characteristic feature of the spin glass systems~\cite{CuDo}, \cite{CrHoSo} and \cite{HeShvN} mentioned in the preceding section: At short times, correlations do not reveal anything about the age of the system.  The data here are rather noisy, but we estimate that $\Delta \propto \tau^{\alpha}$ with $\alpha \approx 0.7$ for all three values of $w$.  We do not see any evidence that $\alpha$ depends on $w$.  

The second noteworthy feature is the tendency to a collapse onto a single function of $\tau/\tau_{w}$ at long times.  This is seen most clearly for the underdetermined case $w=30$ (panels (b) and (c)). This behavior is consistent with the aging behavior found in a wide class of spin glass models~\cite{CK}.  More specifically, the  theoretical expectation from the models described in Appendix A is that $\Delta$ should be proportional to a power of $\tau/\tau_w$ for small values of  $\tau/\tau_w$, and the curves in panel (c) are in fact compatible with this prediction for $\tau$ less than about $6\tau_w$.  The discrepancy at very small $\tau/\tau_w$ can be attributed to the fact that the theoretical prediction is only for the large-$\tau_w$ limit.  For very large $\tau/\tau_w$, Cugliandolo and Le Doussal~\cite{CuDo} found, for their model of a particle diffusing in a correlated random potential in infinite dimensions, that $\Delta$ should grow like $\log(\tau/\tau_w)$.   And, in fact, we find that the behaviour of $\Delta$ for the one case ($\tau_w=80$) where we have measured it for large $\tau/\tau_w$ is consistent with these two limits, as shown in Fig.~10.

For the near-critical $w=68$ (middle row of panels) the behavior is similar to that in the underparametrized case.  Qualitatively, aging is apparently present in the plot (d) as a function of $\tau$ (the curves shift to the right as $\tau_w$ increases).  It is difficult to see any breakdown of the collapse onto a single function of $\tau/\tau_{w}$ in the loglog plot (e), except for the longest $\tau_w$,  $1280$.  However, one can see the breakdown more clearly in the linear plot (f), where the $\tau_w = 1280$ curve stands out from the rest.  

For the overparametrized case $w=90$, the loglog plot (g) as a function of $\tau$ again shows apparent aging at the shorter waiting times, but in the loglog plot (h) as a function of $\tau/\tau_w$ one can see failure of the collapse onto a single function, and this is even more clearly evident in the linear plot (i), especially at the two longest waiting times, $\tau_w = 640$ and $1280$.  

This pattern, with the longest waiting-time $\Delta$ curves beginning to fall below the shorter-waiting-time ones, was also present in the data of Baity-Jesi {\em{et al}}, who identified it as marking the end of aging (see, e.g., their Fig.~4).  Their network was more strongly overdetermined than ours, but here we have shown that the long-time breakdown of aging begins already at the critical width $w \approx 68$.  As in  their data, this interruption of aging occurs around a value of $\tau_w$ equal to the $\tau$  ($\approx 1000$),  at which the loss makes its final approach toward its asymptotic value (see Fig. 3).  

To summarize:  In the underdetermined phase ($w=30$), we find aging behavior something like that in a class of spin glass models, with a collapse onto a single function of $\tau/\tau_{\rm w}$ for $\tau$ of order $\tau_{\rm w}$ or greater.  By analogy with these problems, we might expect aging to disappear at and above the critical $w\approx 68$.  However, even well into the overdetermined phase ($w=90$), aging holds at least approximately at shorter values of $\tau_{w}$, breaking down only for $\tau_w$ of the order of the time it takes the network to reach its best solutions.  

\begin{figure}
 \centering
  \includegraphics[width=0.48\textwidth]{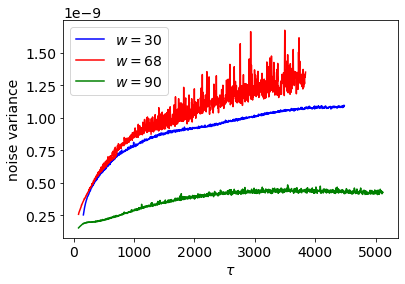}
\caption{Variation of the SGD noise variance with time.}
\end{figure}
 
We have also measured the SGD noise $\Gamma(\tau)$ in our aging runs.  Fig.~11 shows how it varies under learning for $w =30$, $68$ and $90$.  We note a general tendency for $\Gamma$ to grow with $\tau$, but with a decreasing rate of growth and (at least for $w=30$ and $90$) an eventual levelling off.  For the overparametrized case $w=90$, it even appears to fall a bit at the longest measured times.  It is also interesting that the critical case $w=68$ is the noisiest, and the noise variance value varies quite noisily itself.

\section{Discussion}

Our main findings are 

(1) The asymptotic power-law behavior of the loss, 

(2) the critical slowing down that we see upon approaching the transition from the overparameterized side, 

(3) the apparent $\tau_w$-independent growth of the mean-square weight fluctuations $\Delta(\tau_{w}, \tau_{w} + \tau)$ for $\tau \ll \tau_w$,

(4) the finding of aging, specifically the dependence of the normalized growth of fluctuations $\Delta(\tau_{w}, \tau_{w} + \tau)$ only on the ratio $\tau/\tau_{w}$ in the underparametrized phase, and 

(5) the persistence of aging in the overparametrized phase for $\tau_w$ less than the time it takes the loss to reach a value of the order of its minimum possible value. 

The second of these suggests that the interpolation transition may be like the one that occurs at the boundary of spin glass phases in many models, including those of refs~\cite{CuDo}, \cite{CrHoSo} and \cite{HeShvN}.  The third and fourth suggest that underparameterized networks are in a spin glass phase like that found in those models, all of which show this kind of aging.  Indeed, in particular, the model of ref.~\cite{CuDo}, seems like a natural abstraction of our problem.  In it, the components of the position of the particle interact with each other as it moves under the influence of the random potential.  They thus play the role of the weights in our network, and the random potential plays that of the training data. From this perspective, it is not surprising to find the same kind of dynamics in the two systems.

On the other hand, because of the first and last findings in this list, there is apparently a problem in this picture, in which the overparametrized phase should correspond to a normal phase in the spin glass model.  Such a normal phase should exhibit simple relaxational dynamics and no aging, while we find power-law decays of the loss and aging persisting throughout the entire learning process.  We believe the resolution of this  apparent paradox lies in the fact that the theory is constructed only for the large waiting-time limit, that is, for $\tau_w$ much larger than any characteristic relaxation times in the system.  In our networks, this limit is reached only when learning is finished, and this is the condition we see in Fig.~9f for $\tau_w = 1280$ and Fig.~9i for $\tau_w = 640$ and $1280$. In these cases, we see interrupted aging.  Thus, there is no inconsistency between our findings and the dynamics of spin glass models in the limit where they are solved.  Of course this picture does not explain why it takes so long to reach the final equilibrium state and why the equilibration process should exhibit power-law decay, but that problem lies outside the scope of this analogy.

Our findings are consistent with those of Baiti-Jesi {\em{et al}} for non-recurrent layered networks.  Furthermore, we reach the same conclusion as theirs: at truly long times, when the learning is essentially finished, overparametrized networks are not glassy.  Thus, the introduction of recurrent dynamics in out networks does not appear to lead to qualitative changes in the learning dynamics, though it leads to longer learning times.   

In their work, Kerr Winter and Janssen~\cite{WJ} also found dynamical behaviour similar to what have observed.  One feature they  found is something like the $\tau_w$-independent growth of $\Delta(\tau_w,\tau_w+\tau)$ that we see at short times in the first column of Fig.~9.  They call this ``caging", a term used in the theory of structural glasses to describe a state in which atoms or molecules are strongly restricted in their motion by the other molecules around them.   They also observed aging in the underparametrized phase, showing this in sets of weight decorrelation curves that shifted with waiting time, qualitatively similar to what we see in our graphs at longer times.  They do not find this in the overparametrized phase, however.  It would be a complicated task to make more quantitative comparisons and sort out fully the relation between their results and ours. However, it is interesting to find some limited degree of universality of these phenomena.

We should also note a problem we have encountered and not yet dealt with satisfactorily.  We have found it necessary to use very small learning rates (at least an order of magnitude smaller than for networks without recurrent dynamics) to get smooth learning.  This effect gets stronger with increasing network depth and has prevented us from any systematic study of networks with more than 5 layers. It is also exacerbated near the transition, slowing the computation down even more than the critical slowing down alone would imply.   

We have also been limited in what we could do here by the long times required to perform these learning runs.  This limited our aging investigations to 2-layer networks and forced us to rely on single runs and single weight initializations.   However, in the limited investigations of these kinds of variability that we have been able to do, we have not found any hint that the phenomena we have identified are not robust. 

Variations of our model, such as changing the activation function, relaxation of the orthogonality constraint on the recurrent interactions, or replacing the basic layer structure by one using attention mechanisms are also obvious topics that deserve study.  We think the features we have found may be quite general consequences of the nonlinear learning process.  However, exploring this hypothesis would take us far beyond the scope of the present work.  Nevertheless, we think our findings could stimulate further developments in such directions.
 
Finally, we would like to remark that exploring phase diagrams in hyperparameter space and the transitions between phases could be a useful general strategy in complex networks.   It could help in understanding in which regions networks perform stable computations free of undesirable fluctuations and biases, and in which ones they do not, thereby improving the usefulness and reliability of AI systems.

\appendix

\section{Spin glass models}

Thinking of readers who may not be familiar with spin glasses, we sketch here some properties of models that we refer to in the main text.  They are ``infinite range'' models; i.e., there is no notion of distance; every ``spin'' variable interacts with all others, and this feature makes them soluble.

The simplest model of randomly-interacting binary variables is the Sherrington-Kirkpatrick model~\cite{SK}, with energy
\begin{equation}
E = -{\mbox{$\frac{1}{2}$}} \sum_{ij}J_{ij}s_i s_j - \sum_i h_i s_i,					\label{SKmodel}
\end{equation}
where the spins $s_i$ can take values $\pm 1$ and the interactions are iid with zero mean and variance $J^2/N$, where $N$ is the number of spins and $h_i$ is an external field.  For constant $h_i=h$, this model has a spin glass phase below a field-dependent critical temperature $T_g(h)$.  The theoretical description of this phase is highly non-trivial~\cite{Parisi}
and we will not discuss it.   The line separating the normal and spin glass phases is called the AT line, after de Almeida and Thouless~\cite{AT}, who first pointed out its existence.  As one approaches it from the normal side ($T>T_g(h)$), the dynamics exhibit critical slowing down, as described in the Introduction.

Another class of models that is important in our discussions is so-called ``p-spin" models.  In these, the interaction is not generally quadratic as in Eq.~(\ref{SKmodel}), but can involve products of $p$ different spins for $p>2$.  The energy can be written (in zero external field) as
\begin{equation}
E = -\sum_{\{ i_1 i_2 \cdots i_p\}} 
J_{i_1 i_2 \cdots i_p}  s_{i_1} s_{i_2} \cdots s_{i_p} ,      \label{pspinmodel}
\end{equation}
This problem is very hard for binary variables, so it has been studied mostly with continuous $s_i$ subject to the ``spherical constraint'' $\sum_is_i^2 = N$.  This model proves to have dynamics in the spin glass phase which can be fully understood in terms of the kind of aging described in the Introduction.

This model does not have an AT line, but if we add an external field to the problem~\cite{CrHoSo} there is such a line in the $T$-$h$ plane for large enough field.  Furthermore, adding, instead, a simple ferromagnetic interaction $-{\mbox{$\frac{1}{2}$}}J_0(\sum_i s_i)^2$ to the energy~\cite{HeShvN}, generates, for large enough $J_0$, a ferromagnetic state with an internal field that produces the same effect. 

Finally, there is the model studied by Cugliandolo and Le Doussal~\cite{CuDo} of a particle diffusing in a correlated random potential landscape in a space of very high dimensionality.  While this doesn't immediately sound like a spin glass, the kinetic equations for this model are very similar in form to those of p-spin models. They can, for the right kind of correlations in the random landscape, also have an AT line, its associated critical slowing down, and aging like that in p-spin glasses.

\section{Data}
The data were downloaded from the Bengio lab in Montreal, following the link \footnote{www-etud.iro.montreal.ca/~boulanni/icml2012.  This link is apparently no longer active; however, the files are available at github.com/czhuang/JSB-Chorales-dataset.} in their 2012 ICML paper~\cite{BL}.  In this work, we used the first 80 chorales of their collection.  In their data, the music was quantized for simplicity into quarter notes, so half notes and full notes were represented by 2 or 4 repeated quarter notes.  Despite this brutal simplification, one can still recognize the sound of Bach when listening to the music in this form.  

In a preliminary calculation, the chorales were converted into sequences of 54-dimensional binary-valued vectors.  Each direction in this space corresponded to a key on a piano keyboard (54-dimensional rather than 88-dimensional because modern pianos didn't yet exist in Bach's time and he only used 54 notes in these compositions).  For each component, a value $S_i=+1$ indicated that the corresponding key should be played, and a value $S_i=-1$ indicated that it shouldn't be.  Most of the music consisted of 4-note chords (for 4-part chorale singing), so in most of the data vectors, 4 of the components $S_i$ were equal to $+1$, but there were some chords in which some of the voices were silent, so fewer of the components were $+1$.   For each chorale, the training input vectors consisted of all the chords $S_i(t)$ in the sequence they were played in except the last one, and the target output for each input was the chord $S_i(t+1)$ at the following step.  The full 80-chorale training set consisted of 4597 such input-target pairs.

\section{Learning algorithm}

As described in the main text, the dynamics of the network are given by Eqs.~(\ref{hidden0}-\ref{dynamics}) together with the stochastic dependence of the output units $\sigma_i$  on $h_i$.  As the loss function, we took the negative log likelihood
\begin{equation}
L = - \sum_{it} [h_i(t)S_i(t+1) - \log 2 \cosh h_i(t) ]    					\label{lossfn}
\end{equation}
of the training data sequence under the model as the loss function for the learning.  (Equivalently, we could forget the final stochastic step, take the outputs to be the $h_i(t)$ and just say we were using the cross-entropy loss function.) 

The weight change rules for stochastic gradient (SGD) learning can then be obtained by straightforward differentiation.  For $\sf W$, $\sf K$ and $h_0$ they involve the output field components
\begin{equation}
h_i(t) = \sum_a K_{ia} \mu_{d-1}(t) + \sum_j W_{ij} S_j(t) +h_{0,i}
\end{equation}
directly in the target error
\begin{equation}
\epsilon_i(t) = S_i(t+1)-\tanh h_i(t).							\label{targeterror}
\end{equation}
The story for the quadratic hinge loss function
\begin{equation}
L = {\mbox{$\frac{1}{2}$}}\sum_{it} (S_i(t+1) - h_i(t))^2 \Theta(1 -h_i(t)S_i(t))        \label{hinge}
\end{equation}
is almost the same.  The only difference is that the output error  in Eq.~(\ref{targeterror}) is replaced by
\begin{equation}
\epsilon_i(t) = [S_i(t+1)- h_i(t)]\Theta(1 -h_i(t)S_i(t))			\label{hingetargeterror}
\end{equation}
in all the following equations.  Also, when we did hinge loss learning, we also did full (not stochastic) gradient descent, so in that case the minibatch size $p$ in these equations has to be replaced by the full data length.

Thus, for learning $\sf W$, $\sf K$ and ${\vec h}_0$, which connect directly to the output units, we have simply
\begin{equation}
\Delta W_{ij} = \frac{\eta}{p} \sum_{t \in B} \epsilon_i(t)S_j(t)                
\end{equation}
\begin{equation}
\Delta K_{ia} = \frac{\eta}{p} \sum_{t \in B} \epsilon_i(t)\mu_{d-1,a}(t)   
\end{equation}
\begin{equation}
\Delta h_{0i} = \frac{\eta}{p} \sum_{t \in B} \epsilon_i(t)                         
\end{equation}
Here, $B$ denotes the current minibatch and $p$ the minibatch size, $\eta$ is the learning rate, and $\mu_{d-1,a}(t)$ is the activation of hidden unit $a$ in the final hidden layer $d-1$ at time $t$.   

There is a difference between this problem and a standard SGD computation for a network without recurrence.  Here, the forward pass through the layers at time $t$ in some chorale requires (see Eqns. (1-3)) the hidden unit values activations at time $t-1$.  However, these are probably not in the current minibatch of randomly chosen steps, so we don't have access to this information.  Furthermore, in order to know these activations, the ones at $t-2$ are required, and so on all the way back to the beginning of the present chorale.  This means that in order to do SGD in this problem we must, before each learning step, run a full forward pass through all time steps and keep track of the $\mu_{l,a}(t)$ at all $l$ and $t$.  

The remaining learning rules, for the ${\sf J}_l$ and ${\sf M}_l$, can be written simply in terms of effective target errors on hidden units, $\delta_{l,a}(t)$. 
\begin{equation}
\Delta J_{0,ai} = \frac{\eta}{p} \sum_{t \in B} \delta_{0,a}(t)S_i(t)                           
\end{equation}
\begin{equation}
\Delta J_{l,ab} = \frac{\eta}{p} \sum_{t \in B}  \delta_{l,a}(t)\mu_{l-1,b}(t)
\end{equation}
for  $1 \le l \le d-1$, and        
\begin{equation}
\Delta M_{l,ab} = \frac{\eta}{p} \sum_{t \in B} \delta_{l,a}(t)\mu_{l,b}(t-1)
\end{equation}
for $0 \le l \le d-1$.	    
The effective target errors can be computed, starting from the target output errors, by backpropagation through layers and time steps, as we now describe.
 
First, the error backpropagation through the net at every time $t$ starts with the first step
\begin{widetext}
\begin{equation}
\delta_{d-1,a}(t) = X_{d-1,a}(t)[ \sum_i \epsilon_i(t) K_{ia} + \sum_b \delta_{d-1,b}(t+1)M_{d-1,ba}]  
\end{equation}   
\end{widetext}
with 
\begin{equation}
X_{l,a}(t) = 1 - \mu_{l,a}^2(t)
\end{equation}
(the derivative of the activation function).  The first term describes the backpropagation at time $t$ from output errors at units $i$, through the $\sf K$ matrix to hidden unit $a$ in the last hidden layer.  The second term describes backpropagation in time (BPTT) of effective errors on unit $b$ in the last hidden layer, via the ${\sf M}_{d-1}$ matrix, to unit $a$ in that same layer one step later.  Note that the first term doesn't change $t$ (propagation forward and back in the network is, by construction, instantaneous), and the second term doesn't change the layer index (the $\sf M$ matrices describe intralayer interactions).  In the program, these calculations for different $t$ are effectively simultaneous (done in parallel).

Now it is simple to extend this, recursively, to layers with smaller $l$:
\begin{widetext}
\begin{equation}
\delta_{l,a}(t) = X_{l,a}(t)[\sum_b \delta_{l+1,b}(t) J_{l+1,ba}  +\sum_b\delta_{l,b}(t+1) M_{l,ba}]
\end{equation}
\end{widetext}

In both Eqns. (B8) and (B10), we need to know the effective target on units in the same layer one time step later than $t$.  In a conventional BPTT calculation with only a single layer, one conventionally works back iteratively one step at a time in the same way we worked back in through layers in the above equations.  Here, instead, taking advantage of the fact that we have calculated all the $\delta_{l,b}(t)$  in the previous learning step, we can simple shift that array of $\delta$'s by one time step and use that on the right hand side of Eqns. (B8) and (B10).   We also found it advantageous to smooth this process by using the average of the $\delta$'s from the previous two steps, rather than just those from the most recent step.  In either case, the result is not the exactl solution of these equations, but if the learning is converging to a stationary solution, it can work.  In practice, we find satisfactory convergence for slow enough learning rates.

In both the forward and back-propagation parts of the learning algorithm, it is necessary to prevent information to cross boundaries between successive chorales in the training example sequence.  We do this by, at each learning step, setting the hidden unit activations $\mu_{l,a}(t)$ and the effective target errors $\delta_{l,b}(t)$ to zero whenever $t$ is the first chord in a chorale.

As described in the main text, the networks were run using this kind of BPTT and SGD with a minibatch size $p=300$.   The cost function was the negative log likelihood of the data under the model.  Exceptionally, in the case of the learning runs described in Section 4 for two hidden layers, full gradient descent was used.  However, SGD was used in the aging calculations for $d=2$ and the learning in them showed no significant difference from the full gradient descent calculations.  

Both the feedforward matrices ${\sf J}_a$ (except the initial input matrix ${\sf J}_0$) and the intralayer recurrent matrices ${\sf M}_a$ were initialized to be orthogonal.  ${\sf J}_0$ was initialized as random with elements of variance $0.1/ N_v$, with $N_v$ the input dimensionality.  The output matrices $\sf K$ and $\sf W$ were initialized to zeros. We haven't found it necessary to include biases in the inputs to any units except the outputs, presumably because we rescaled our inputs to have zero mean and all the connection matrix elements are also zero-mean.

The orthogonality of the ${\sf M}_a$ was enforced throughout the learning by the following procedure:  After the weight changes were made at each learning step, a singular value decomposition of each matrix was made and all the singular values were set to 1.  

Initial learning rates ranged from $10^{-3}$ for single-layer networks down to $10^{-4}$ for 10-layer networks.  For $d=5$ and $10$, we discovered that if such rates are maintained, the NLL would show chaotic spiky fluctuations well before the target value (usually $0.01$) was reached.  Therefore, learning was usually done for $2 \times 10^5$ iterations at a time, and the learning rate was lowered, typically by a factor of $2$, whenever these fluctuations started to occur.  Final learning rates were as low as $2.5 \times 10^{-5}$ for $d=5$ and $1.25 \times 10^{-5}$ for $d=10$ (in both cases for $w \le 60$).  Typical learning runs required $2$ to $10$ million iterations and took 40-150 hours on MacBook Pros with an M1 or M3 processor.

\section{Test losses}

We performed a few long runs for 2-hidden-layer networks in which we recorded the loss on a test set based on 80 chorales not used in training.  The training data for these runs were those used in making Fig.~6. The width values are 35 (about half the critical width), 70 (very near the critical width) and 140 (twice the critical width). Fig.~12 shows the results of runs of 1.5 million learning steps.  

\begin{figure}[h!]
 \centering
  \includegraphics[width=0.48\textwidth]{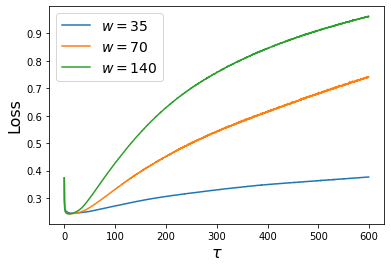}
\caption{Evolution of the test loss under training}
\end{figure}

These data do not show the feature, observed in some systems~\cite{Spigleretal,Belkinetal,MeiMontanari}, of a maximum in the test loss around the interpolation transition.  This could be a consequence of the small size of our training set (4973 chord transitions).  We were limited to such a training set size by the long learning times we had to explore.  Alternatively, if we could make these runs for an order of magnitude longer times, the curve for $w=70$ might cross that for $w=140$.  However, this question is not within the focus of our investigation here, so we have not explored this possibility.

\bibliographystyle{apsrev4-1}
\bibliography{nets}

\end{document}